\documentstyle[epsf,colordvi,12pt]{article}

\begin{document}

\title{\bf{The Short-Time Critical Behaviour of the Ginzburg-Landau Model
with Long-Range Interaction}
\thanks{ Supported by the National Natural Science
Foundation of China under the project 19772074 and 
by the Deutsche Forschungsgemeinschaft under the
project Schu 95/9-2. }}
\author{\normalsize \sf Y. Chen $^{a}$, S. H. Guo$^{a}$,
Z. B. Li$^{a,b}$,  S. Marculescu$^{c}$ and L. Sch\"{u}lke$^{c}$ \\
\normalsize $^{a}$ Zhongshan University,
Guangzhou 510275, China \\
\normalsize $^{b}$ Associate Member of ICTP, Trieste, Italy \\
\normalsize $^{c}$ Universit\"{a}t-GH Siegen, D-57068 Siegen,
Germany } 
\date{}
\maketitle
\vskip 1.5pc

\begin{abstract}
The renormalisation group approach is applied to the study of
the short-time critical behaviour of the $d$-dimensional Ginz\-burg-Landau
model with long-range interaction of the form $p^{\sigma} s_{p}s_{-p}$ in 
momentum space. Firstly the system is quenched from a high temperature to
 the critical temperature  and then relaxes to equilibrium within the 
model A dynamics. The asymptotic scaling laws and the
initial slip exponents $\theta^{\prime }$ and $\theta$ of
the order parameter and the response function respectively, 
 are calculated to the second order in $\epsilon=2\sigma-d$.
 

\end{abstract}

\newpage
\section*{1. Introduction}

In recent years, much attention has been paid to the 
short-time critical dynamics. 
The short-time phenomena arise at 
times just after a microscopic time scale $t_{mic}$
needed by the system to remember only the macroscopic condition and to forget
all specific microscopic details. 
The corresponding time regime is also called critical initial slip
in order to 
distinguish it from the uninteresting microscopic time interval 
between zero and $t_{mic}$.
Since the pioneering analytical study of  \cite{s1}, universal 
short-time scaling has been found in various models ( See \cite{s2} and 
\cite{luo}). 
When the system is quenched from a high temperature $T_i$ to the
critical   temperature $T_{c} \ll T_i$  the order parameter shows in the
short-time regime a power law increase $m(t) \sim  t^{\theta^{\prime}}$ with a
new universal critical exponent $\theta^{\prime}$. 

The short-time dynamics has been thoroughly investigated for models with
short-range interaction (SRI). Since the critical equilibrium properties
are modified by the presence of long-range interactions (LRI)
it may be interesting to know how the short-time critical behaviour depends
upon the interaction range. 

The statistical mechanics of LRI has a long history. Already in the 60's
Jovce \cite{jov66} studied thermodynamic properties of the static 
spherical model
with long-range ferromagnetic interaction between the spins. 
The static critical exponents for LRI have been computed 
for the $n$-vector model
by use of the renormalisation group approach [4-8] 
and the ${1/n}$-expansion techniques \cite{suz73}. There are also Monte Carlo
simulations for the one-dimensional static model \cite{lui97}. 

The dynamic properties of the LRI in the long-time regime 
have also been studied as early as the 70's.
Suzuki et al. \cite{ suz75} extended the dynamic theory 
developed by Halperin et al. \cite{hal72} to investigate
an exponent which describes the critical slowing
down in  the $n$-vector model 
with LRI for $T\ge T_{c}$, with equilibrium initial conditions. Folk and Moser
studied a three-dimensional dynamical model for liquids and demonstrated that
the  critical dynamics was affected by LRI \cite{fm94}. The kinetic spherical
model  showed that the short-time critical exponents were  modified by LRI
\cite{chen99}. 

In this paper, we study the short-time critical behaviour of  
the dynamic Ginzburg-Landau model with long-range
exchange interaction. 
In equilibrium at temperature $T$ the O($n$) symmetric Hamiltonian is given by
   \begin{equation}
   \label{hami}
   H[s] \equiv \int d^{d}x \left\{ {a\over 2}(\bigtriangledown s)^{2}+
   {b\over 2}(\bigtriangledown^{
   {\sigma \over 2}}  s)^{2}+{\tau \over
   2}s^{2}+{g \over 4!}(s^{2})^{2} \right\} \;
   \end{equation}
where $s=(s^{\alpha})$ are $n$-component spin fields, $\tau$ is proportional 
to the reduced temperature $1 - T/T_{c}$ and $g$ is the coupling constant. 
The SRI model corresponds to $a=1$ and $b=0$, whereas for the pure LRI model 
$\sigma < 2$, $a=0$ and $b=1$. Since the case $0<\sigma <d/2$ is covered
by a mean-field theoretic description, and since for $\sigma > 2 $ and $ d >
2$ the model (1) belongs to the same universality class
as the SRI model, we will restrict ourselves in the present paper to the range
$d/2 <\sigma < \min (2, d)$.

The  dynamics to be discussed here 
is controlled by the Langevin equation

\[ \partial_{t} s^{\alpha}(x,t) 
   =-\lambda \; {\delta H[s] \over \delta s^{\alpha}(x,t)}+{\xi}^{\alpha}(x,t)
 \; \]  
where $\lambda$ is the kinetic coefficient.
The random forces $\xi =(\xi^{\alpha})$ are assumed to be 
Gaussian distributed
\[  < \xi^{\alpha} (x,t) >=0 \; ; \qquad  \; <\xi^{\alpha} 
   (x,t)\xi^{\beta} (x',t') >=2\lambda
   \delta^{\alpha \beta} \delta(x-x')\delta (t-t') \; . \]

As mentioned above, the initial state is prepared
(macroscopically) at some very high temperature $T_i$. One assumes that the
initial condition $s_0(x) \equiv s(x, 0)$ has also a Gaussian distribution
$P[s_{0}]\propto \exp(-H^{i}[s_{0}])$ where
\[ H^{i}[s_{0}] \equiv \int d^{d} x {\tau_{0} \over
2} {\left[ s_{0}(x)-m_{0}(x) \right]}^{2} \; ,	\]
${\tau}_0$ is proportional to $1 - T_i/T$
and $m_{0}(x)$ is the (spatially
varying) initial order parameter. Being away from criticality ($T_i \gg T_c$),
  the initial correlation function will be short-ranged. 
Since $\tau_{0} 
\sim \mu^{\sigma}$ (where $\mu$ is a renormalisation
momentum scale), the physically interesting fixed point is
$\tau_{0}^{*}=+\infty$, which
corresponds to a sharply prepared initial 
state with initial order $m_{0}$ and zero correlation length.

Introducing
a (purely imaginary) response field
$\widetilde s(x,t)$ \cite{msr73,dom76}, the generating
functional for all connected correlation and  
response functions is given by
   \begin{equation}
   W[h,\widetilde h]=\ln \int {\cal D}(i\widetilde s,s) \exp 
   \left\{ -{\cal L} [\widetilde s,s]
   -H^{i}[s_{0}]+\int\limits_{0}^{\infty}dt \int d^{d}x (hs+\widetilde h
   \widetilde s) \right\} \;
   \end{equation}
where 
   \begin{equation}
   \label{larg}
   {\cal L} [\widetilde s,s]\equiv \int\limits_{0}^{\infty} dt \int d^{d}x 
   \left\{ \widetilde
   s \left[ \dot{s} +\lambda \left( \tau-a\bigtriangledown^{2}+b
   (-\bigtriangledown^{2})^{{\sigma \over 2}} \right ) s
   +{\lambda g \over 6}ss^{2} \right] -\lambda\widetilde s^{2} \right\} 
   \; .
   \end{equation}
Here we have used a pre-point discretisation with respect to time so that
the step function $\Theta(t=0)=0$. Then the contribution
(proportional to $\Theta(0)$) to $ {\cal L} [\widetilde s,s]$ arising from 
the
functional determinant $\det \left[ \displaystyle{\frac{\delta \xi (x,
t)}{\delta s(x,t)}} \right]$ vanishes \cite{di87}.

It is believed that the singularity of the temporal 
correlation is essential to the short-time scaling and 
the scaling can emerge in the early stage of the evolution even though
all spatial correlations are still short-ranged.

The system is now rapidly quenched to a temperature $T \simeq T_c$. The order
parameter will undergo a relaxation process displaying an initial
increase. As long as the correlations are short-ranged and the spatial
dimension $d$ is smaller than the critical dimension $d_c$, the order
parameter follows a mean-field ordering process because the mean-field
critical temperature $T_c^{({\rm m} {\rm f})}$ is larger than the actual
critical temperature $T_c$. This ordering causes an amplification of the
initial order parameter. For $d > d_c$ mean-field theory applies and there is
no critical increase.  

For the SRI models $d_{c}=4$.  The 
longer is the interaction range, the stronger the  suppression 
of  the fluctuations and hence the critical dimension of the LRI
model is smaller. Indeed, it turns out that $d_c = 2 \sigma$. 
Also one would expect that the critical initial increase 
should be weaker as the interaction range becomes longer. 

Since the short-range exchange interaction is irrelevant for $d/2 < \sigma <
\sigma_{s} \equiv 2-\eta_{sr}$ where $\eta_{sr}$ is the  Fisher exponent 
at the
SRI fixed point, one can
consider only pure LRI. We apply the $\epsilon$-expansion theory to the
LRI model in this regime  with $\epsilon \equiv 2 \sigma - d$. The critical
initial order increase  appears in the LRI model for $1 \le d <d_{c}$. 
The scaling behaviour of the critical initial slip is governed by the
exponents $\theta$ and ${\theta}'$. They are computed as functions of $d$
and $\sigma$.

 For $\sigma$ close to (but larger
than) $d/2$ the quantity $\epsilon$ is very small and
the numerical values of the exponents are accurate when computed to 
order ${\epsilon}^2$.

However, when the interaction range is  not very long, 
the situation becomes more complicated, due to a subtle competition between 
the SRI  and the LRI fixed points [6-9]. 
Honkonen \cite{hon90} 
computed the $\beta$-function of the renormalisation group for the pure LRI
model at three-loops and found that the infrared LRI fixed point becomes
unstable for $\sigma = {\sigma}_s$. In the pure LRI model the exchange
interaction term is not renormalised, so that the anomalous dimension of the
field $s(x,t)$ is zero, whereas in the SRI model the field carries some
anomalous dimension $\gamma$. Taking the limit $\sigma \rightarrow 2$ the
expressions for the anomalous dimension (and for other critical exponents) do
not coincide. 
However, as first shown by Sak \cite{sak73} to the leading non-trivial
order and later by Honkonen and Nalimov \cite{hn89} to all order in
${\epsilon}' \equiv 4 - d$, the anomalous dimension $\gamma$ and the other
exponents are continuous functions of the parameter $\sigma$. This means
that the scaling regime of the LRI model is valid only for $\sigma <
{\sigma}_s$, whereas for $\sigma > {\sigma}_s$ the scaling behaviour is
described by the SRI model. At the borderline value $\sigma  = {\sigma}_s$ the
two descriptions yield equal values for the critical exponents.
Let us conclude here by remarking that these last results were obtained solely
for static models.  

The  paper is organised as follows: 
In section 2,
the LRI model with $\sigma<\sigma_{s}$
is studied by the $\epsilon$ expansion method.
The scaling behaviour of the order parameter, correlation and
response functions, as well as the corresponding critical 
initial slip exponents, are obtained.
Section 3 contains conclusions and discussions.

\section*{2. The short-time scalings and exponents}

Since the SRI is irrelevant for $\sigma<\sigma_{s}$, in this section
we take $a=0$ and $b=1$ in (\ref{larg}).

For $g=0$, the generating functional (2) becomes  Gaussian 
and can be easily evaluated in momentum space. One must take into account 
the initial condition, by imposing the following boundary conditions:
\[
\widetilde{s} (x, \infty ) = 0 \; \qquad \qquad s_0(x) = m_0(x) + 
{{\tau}_0}^{-1} \widetilde{s}(x, 0) \; .
\]

The free response function $G_{p}(t,t')$ and the free 
correlation function $C_{p}(t,t')$ are respectively
\begin{eqnarray}
G_{p}(t,t') & = & \Theta (t-t')\exp[-\lambda (p^{\sigma}+\tau)(t-t')] \; ;
 \nonumber \\ 
C_{p}(t,t') & = & C_{p}^{(e)}(t-t')+C_{p}^{(i)}(t,t') \; , \nonumber
\end{eqnarray}
with equilibrium part $C_{p}^{(e)}(t-t')$ 
and initial part $C_{p}^{(i)}(t,t')$ defined by
 \begin{eqnarray*}
 C_{p}^{(e)}(t-t')& \equiv & {1 \over \tau + p^{\sigma}} \exp [-\lambda 
 (p^{\sigma}+\tau )|t-t'|] \; ;  \\
 C_{p}^{(i)}(t,t')& \equiv &\left ( \tau_{0}^{-1}
  -{1 \over \tau + p^{\sigma}} \right ) \exp [-\lambda 
 (p^{\sigma}+\tau )(t+t')] \; .
 \end{eqnarray*}

One sets now a perturbation expansion ordered by the number of loops in the 
Feynman diagrams. It is convenient to consider the Dirichlet boundary
conditions ${\tau}_0 = + \infty$ and $m_0(x) = 0$. The general case is
recovered by treating the parameters ${{\tau}_0}^{-1}$ and $m_0(x)$ as
additional perturbations.

The model (2) with Dirichlet boundary conditions must be renormalised. For this
purpose notice that the free correlation function simplifies to
\[
C_p^{(D)} (t,t')\equiv {1 \over \tau + p^{\sigma}} \left\{ \exp [ - \lambda
(p^{\sigma} +  \tau )|t - t'|] - \exp[ - \lambda (p^{\sigma} + \tau )(t + t')]
\right\} \; .
\]
By integrating over the internal momentum and time coordinates one encounters
ultraviolet divergences which can be absorbed through the reparameterization
of a finite number of coupling constants and fields.
 
Through dimensional analysis, one can show that the critical dimension 
$d_{c}=2 \sigma$ and hence it is convenient to make an expansion in 
$\epsilon = 2 \sigma - d$. 
We will adopt the dimensional regularisation  with minimal subtraction  
scheme \cite{ho73} and introduce renormalised quantities through 
multiplicative factors
\begin{eqnarray}
      s_{b} & = & Z_{s}^{1/2} s \; , \qquad \; {\widetilde s}_{b}=
Z_{\widetilde s}^{1/2}\widetilde s \; ,\qquad \; {\lambda}_{b}=
(Z_{s}/ Z_{\widetilde s})^{1/2} \lambda \; , \nonumber \\
 {\tau}_{b} & = & Z_{s}^{-1} Z_{\tau} \tau \; , \qquad \;
   g_{b} = K_{d}^{-1}{\mu}^{\epsilon} Z_{s}^{-2} Z_{u} u \; ,  \nonumber \\    
{\tau}_{0b} & = & (Z_{\widetilde s} / Z_{s})^{1/2} {\tau}_{0} \; ,\qquad
\; {\widetilde s}_{0b}=(Z_{\widetilde s} Z_{0})^{1/2}{\widetilde s}_{0}  
\end{eqnarray}
where the subscript $b$ denotes the bare quantity and $ K_{d} \equiv
2^{1-d}{\pi}^{-{d \over 2}} [\Gamma ( d / 2)]^{-1}$. 

Some comments are in order: 

(i) The graphs containing only the equilibrium part of
the correlation function are associated to 1PI diagrams and can be made
finite by the same renormalisation factors as the translation-ally invariant
theory. 

(ii) In addition there are divergences arising for $t + t' = 0$
from the initial part of the
correlation function. Remarkably enough, such divergences 
can be multiplicatively removed if one associates them with $n$-point
connected  Green functions. A simple dimensional analysis reveals that new
re-normalisations are required only in two-point functions. Due to the Ward 
identities:
\[   
s_0(x) = 0 \; \qquad \qquad \dot{s}_{0} (x)=2\lambda \widetilde s_{0}(x) \;    
\]  
which hold when inserted in the connected Green functions, one is left with a
single additional renormalisation constant $Z_0$.

A two-loop calculation gives the following 
renormalisation constants: 
  \begin{eqnarray}
  \label{zs}
  Z_{s} & = & 1   \; ; \\
  Z_{\widetilde s}&=& 1-{n+2 \over 6 \epsilon } B_{\sigma} 
  u^{2} \; ;  \\
  Z_{u}&=& 1+ { n+8 \over 6 \epsilon}u + \left [ {(n+8)^{2} \over 36 \epsilon^{
  2}} -{ 5n+22 \over 36 \epsilon} D_{\sigma}\right ]u^{2} \; ;  \\
  Z_{\tau}&=& 1+ { n+2 \over 6 \epsilon}u + \left [
  {(n+2)(n+5) \over 36 \epsilon^{2}}-{n+2 \over 24\epsilon }D_{\sigma}
  \right ] u^{2} \; ; \\
  Z_{0}&=& 1+ { n+2 \over 6 \epsilon}u + {n+2 \over 12 \epsilon^{2}}
  \left [ {n+5 \over 3}+\left ( {2\over \sigma}\ln 2 - {1\over 2} D_{\sigma}  \right) 
  \epsilon \right ] u^{2} \: . 
  \end{eqnarray} 
Here we have introduced 
\[ B_{\sigma} \equiv K_{2\sigma}^{-1}\int
{d^{2\sigma}x \over (2\pi)^{2\sigma}} [1+x^{\sigma}+( {\bf e}
+ {\bf x})^{\sigma}]^{-2} x^{-\sigma} \; \]
with $\bf e $  a unit vector in the $2\sigma$-dimensional space, and
\[ D_{\sigma} \equiv
\psi (1)-2 \psi (\sigma /2) + \psi (\sigma) \; \]
with $\psi (x)$ the logarithmic derivative of
the gamma function. For the  particular case $\sigma=2$, one has
$B_{2}={1 \over 2} \ln (4/3)$,  and $D_{2}=1$. The calculation
of the renormalisation constant $Z_{0}$ is reported in Appendix.

According to the general solution of the renormalisation group equation, 
the renormalised connected Green function of 
$N$ $s$-fields, $\widetilde N$ $\widetilde s$-fields,
and $M$ $\widetilde s_{0}$-fields at the fixed point $u^{*}$ has the following
scaling law:
\begin{eqnarray}
& & G_{N\widetilde N}^{ M}(\{ x,t \} ,\tau , 
\tau_{0}^{-1},\lambda, u^{*}, \mu) 
= l^{(d-\sigma+\eta_{s}) N / 2 + (d+\sigma+\eta_{\tilde{ s}})
{\widetilde N} / 2 +(d+\sigma +\eta_{\tilde{ s}}+\eta_{0}) 
M / 2} \nonumber \\
& & \mbox{} \times G_{N\widetilde N}^{ M}
(\{ lx, l^{\sigma+\zeta (u^{*})} t \} ,\tau l^{-\sigma+\kappa (u^{*})} , 
\tau_{0}^{-1}l^{\sigma+\zeta (u^{*})},\lambda, u^{*}, \mu ) \;
\end{eqnarray}
where $\eta_{s}\equiv \gamma (u^{*})$, 
$\eta_{\tilde{s}} \equiv \widetilde \gamma (u^{*})$,
and $\eta_{0} \equiv \gamma_{0} (u^{*})$. The Wilson functions entering the
renormalisation group equations are defined by     
\[ \begin{array}{lll}
  \gamma \equiv \mu \partial_{\mu} \ln Z_{s}|_{0} \; ; & \beta \equiv 
   \mu \partial_{\mu} u|_{0} \; ; & \ \widetilde \gamma \equiv  \mu
\partial_{\mu}     \ln Z_{\widetilde s}|_{0}  \; ; \\
   \kappa \equiv \mu \partial_{\mu} \ln \tau|_{0} \; ;  & \zeta \equiv \mu 
   \partial_{\mu} \ln \lambda |_{0}={1 \over 2} ( \widetilde
   \gamma - \gamma)\;  ; & \gamma_{0} \equiv \mu \partial_{\mu} \ln
Z_{0}|_{0} \;   
\end{array} \]
and are computed perturbatively from Eqs.(5-9).
The symbol $|_{0}$ means that $\mu$-derivatives are calculated at fixed
bare parameters. For instance, at the two-loop level, the Wilson function 
$\gamma_{0}$ (related to the initial order parameter) is given by
  \begin{equation}
  \label{eta0}
  \gamma_{0}=-{n+2\over 6} \left [ 1 + \left (
  {2\over \sigma}\ln 2-{1\over 2}
  D_{\sigma} \right ) u \right ] u \; .
  \end{equation}
By solving algebraically the equation $\beta (u) = 0$ one finds the infrared
LRI fixed point
\begin{equation}
u^{*}={6 \epsilon \over n+8} \left [ 1+{2(5n+22) \over  (n+8)^{2}} 
D_{\sigma} \epsilon \right ] + O(\epsilon^{3})
\end{equation}
and subsequently the values of the Wilson functions at this point.

In order to identify the critical exponents one can compare the standard
scaling form of the two-point correlation function 
\[
 G_{20}^{0}( x-x',t, t'; \tau)=|x-x'|^{-(d-2 +\eta)}f \left( {|x-x'| \over
\xi},  {|x-x'| \over t^{1/z}}, {|x-x'| \over {t^{\prime}}^{1/z}} \right) 
\] 
to the Eq.(10) in which we have set $N=2$, 
$\widetilde N = M = 0$ and $lx=1$. Here $\xi \equiv {\tau}^{-\nu}$.

In this way we find the LRI critical exponents to second order in $\epsilon$
\cite{fis72,suz75}
  \begin{eqnarray*}
  \eta & \equiv & 2-\sigma + {\eta}_s = 2 - \sigma \; ; \\
  z & \equiv & \sigma + \zeta (u^{*}) = \sigma + {6(n+2) \over
  (n+8)^{2}} B_{\sigma} \epsilon^{2} \; ; \\
  1/\nu & \equiv & \sigma - \kappa (u^{*}) = \sigma -{n+2 \over n+8}\left [ 1+
  {7n+20 \over (n+8)^{2}} D_{\sigma} \epsilon \right ] 
  \epsilon \; .
  \end{eqnarray*}
Notice that the anomalous dimensions of $s$ and $\widetilde s$ 
are $\eta_{s}=2-\sigma+\eta$ and $\eta_{\widetilde s}=\eta+2(z-\sigma)$
respectively, whereas that of the initial order parameter $\eta_{0}$
is given by Eq.(11) at the fixed point (12).
  
Employing a short-time expansion  of the fields
$s(x,t)$ and $\widetilde s(x,t)$, as done in \cite{s1},
one can derive the following behaviour of the full response 
and  correlation functions for $t > 0$ but $t' \rightarrow 0$:
\begin{eqnarray}
G(p,t,t') & = & p^{-2+\eta+z}
\left ( {t \over t'} \right )^{\theta} f'_{G} 
\left ( p \xi, p^{z}t \right ) \;  \nonumber \\ 
C(p,t,t') & = & p^{-2+\eta} 
\left ( {t \over t'} \right )^{\theta -1} f_{C}
\left ( p \xi,p^{z}t \right ) \; .
\end{eqnarray}
Here we defined the initial slip exponent $\theta$ and computed it to 
second order in $\epsilon$ 
\[
\theta \equiv -{\eta_{0} \over 2z}={\epsilon (n+2) \over 2\sigma (n+8)}
\left\{ 1+\left[ {7 n +20 \over  (n+8)^{2}}D_{\sigma}+{12\ln 2 
\over \sigma (n+8)}
\right] \epsilon
\right\} \; .
\] 

Let us discuss now the scaling form of the order parameter which
relaxes from a  non-zero initial value $m_{0}$ to zero. As mentioned above we 
can consider $m_{0}(x)$ an additional time independent source coupled to the
initial response field $\widetilde s_{0}(x)$.
Owing to the renormalisation of the initial order parameter  
\[ m_{0b}(x)=(Z_{0}Z_{\widetilde s})^{-1/2}m_{0}(x) \; ,\]
no new renormalisation is required for 
the time dependent order parameter $m(x,t) \equiv 
<s(x,t)>|_{\widetilde{h}=h=0}$.  By taking a homogeneous source
$m_{0}(x)=m_{0}$, but keeping still $\tau_{0}^{*}=+\infty$, we obtain the
power law 
\begin{equation}
m(t)=m_{0} t^{\theta^{\prime}} f_{m} \left ( m_{0}t^{
\theta^{\prime}+{d-2+\eta \over 2z}} ,\tau t^{{1 \over \nu z}} \right )
\end{equation}
where the exponent $\theta^{\prime}$ is defined by 
\[
\theta^{\prime} \equiv -{\eta_{s}+\eta_{\tilde{s}}+\eta_{0} 
\over 2z} \; .
\]
To second order in $\epsilon$ it has the value
\[
{\theta}'= {\epsilon (n+2) \over 2\sigma (n+8)}
\left\{ 1+\left[ {7 n+20 \over 
(n+8)^{2}}D_{\sigma}+ {12 \left( \ln 2 
-\sigma B_{\sigma} \right)\over \sigma (n+8)}\right] \epsilon \right\}\; .
\]
As first indicated in \cite{s1} for the SRI model, the critical exponents  
$\theta$ and $\theta^{\prime}$ 
are related by ${\theta}^{'}=\theta+(2-z-\eta)/z$. 

The function $f_{m}(x,y)$ appearing in (14) has a universal behaviour at the
critical point $\tau=0$: $f_{m}(0,0)$ is  finite; while for $x \rightarrow
\infty$, $f_{m}(x,0)$ behaves like $\sim  1/x$.

\section*{ 3. Discussions and conclusions}

When the long range interactions are dominant, 
$\epsilon$ is small enough and the calculated values of 
$\theta^{\prime}$ and $\theta$ for physical dimensions are numerically
reliable. For instance, we list in Table \ref{t1} the values corresponding
to $\epsilon=0.1$ for $n=1$ and $d=1,\ 2, \ 3$. 

\begin{table}[h]\centering

\begin{tabular}{|c|c|c|c|}\hline
  & $d=1,\ \sigma=0.55$ & $d=2, \ \sigma=1.05$ & $d=3, \ \ \sigma=1.55 $ \\ \hline
$\theta^{\prime}$ & 0.0383 & 0.0180  & 0.0117\\ \hline
$\theta$ & 0.0408 & 0.0187 & 0.0120 \\ \hline
\end{tabular}
\caption{ \small{The values of $\theta^{\prime}$, $\theta$ to
$\epsilon=0.1$ for $n=1$ and $d=1,\ 2, \ 3$.}
}
\label{t1}
\end{table}

In the following, we will focus the discussion on ${\theta}'$. Let us first 
notice that both the response and the
correlation functions measure the fluctuations of the spin-fields. Since 
$\theta$ and ${\theta}'$ are positive, one expects, according to Eqs.(13),
an initial increase of the fluctuations. Of course, the increase depends upon
$\sigma$ and $d$.

In Figure 1 
the exponent $\theta^{\prime}$ is plotted versus $d$ for 
$\sigma = 1/2, \ 1, \ 3/2$ and 2 respectively and $n = 1$.  
The value $\sigma=2$ corresponds to the SRI model. 
At fixed $\sigma$, the
exponent ${\theta}'$ decreases when $d$ increases, because fluctuations
are reduced as the dimension gets larger.
At the critical dimension $d_c = 2 \sigma$ the value of $\theta^{\prime}$ 
becomes equal
to zero. Here other scaling laws would replace the power law.

Figure 2 
shows that $\theta^{\prime}$ for $d=2$ and small
$\sigma$ monotonously increases with $n$. For larger $\sigma$ it first
reaches a peak and then it decreases toward $n \rightarrow \infty$. For
other values of the spatial dimension $d$ the pictures are  similar. 
The increase of $\theta^{\prime}$ can be easily understood as  
more internal degrees of freedom (larger $n$) help the fluctuations increase. 
Hence the critical behaviour is smooth in $n$ and can be studied in an 
$1/n$-expansion.
But for large $\sigma$ one can reach the opposite effect
\cite{ma76}, the fluctuations decrease when $n$ exceeds some threshold value.

In Figure 3 
the exponent ${\theta}^{'}$ is plotted versus $\sigma$ for $n=1$ and $d=1, 2
, 3$. At fixed $d$ the exponent ${\theta}'$ decreases when $\sigma$ decreases,
because the fluctuations are more suppressed by interactions of longer range 
($\sigma$ smaller).
In one dimension, there is no
SRI fixed point hence only the curve controlled by the LRI fixed point 
is observed. 
 
All the previous considerations were of qualitative nature since they do not 
take into account the interaction of coupling constant $g$. Of special
interest in this respect is the LRI fixed point (12).
 At 
$\sigma = {\sigma}_s \equiv 2 - {\eta}_{sr}$ (where ${\eta}_{sr} \equiv 
\displaystyle{\frac{n + 2}{2 ( n + 8 )^2}} {{\epsilon}'}^2$ and ${\epsilon}'
\equiv 4 - d$) and fixed $d$ we have
\begin{eqnarray}
\epsilon & = & {\epsilon}' - \frac{n + 2}{4 ( n + 8 )^2} {{\epsilon}'}^2  
\; ; \\
u^{*}    &  = &  \frac{6 {\epsilon}'}{n + 8} \left[ 1 + 
\frac{3 ( 3 n + 14 ) {\epsilon}'}{( n + 8 )^2} \right] \equiv u_{SR}^{*} 
\; ; \nonumber \\
{\theta}' & = & \frac{{\epsilon}' ( n + 2 )}{4 ( n + 8 )} \left[ 1 +  
 + \frac{6 {\epsilon}'}{n + 8}  \left( \frac{n + 3}{n + 8} + \ln
\frac{3}{2} \right) \right]  \equiv {{\theta}'}_{SR} \; . \nonumber
\end{eqnarray}
Here the subscript $SR$ means short-range regime.

In order to explore the limitation of $\sigma \rightarrow 2$ of the LRI, we 
make a double expansion in $\epsilon$ and $\alpha \equiv 1-\sigma/2$ with
$\alpha$ of the order $\epsilon$ or smaller. The infrared fixed point to 
order ${\epsilon}^2$ is located at
\begin{equation}
u_{wlr}^{*} = \frac{6 \epsilon}{n + 8} \left\{ 1 + 
\frac{\epsilon}{( n + 8 )^2} \left[ 3 ( 3 n + 14 ) +  ( n + 2 ) 
\frac{\alpha}{\alpha + \epsilon} \right] \right\} \; . 
\end{equation}
Here the subscript $wlr$ (weak-long-range) means that $\alpha$ is at most of
order $\epsilon$. The critical initial slip exponent in the weak-long-range
limit can be also computed to this order: 
\begin{equation}
{{\theta}'}_{wlr} = \frac{\epsilon ( n + 2 )}{4 ( n + 8 )} \left\{ 1 +  \alpha 
 + \frac{\epsilon}{n + 8} \left[ 6 \left( \frac{n + 3}{n + 8} + \ln
\frac{3}{2} \right) +  \frac{n + 2}{n+8} \frac{\alpha}{\alpha + \epsilon}
\right] \right\} \; . 
\end{equation} 
Since $\alpha$ is actually of order
${\epsilon}^2$ it can be set to zero in (16) and (17), and taking 
into account that 
$\epsilon = {\epsilon}' + O ({\epsilon}^2)$, one gets
\begin{eqnarray}
u_{lwr}^{*}       &  = &  \frac{6}{n + 8} \left[ \epsilon + 
\frac{3 ( 3 n + 14 ) {{\epsilon}'}^2}{( n + 8 )^2} \right]  
\; ; \nonumber \\
{{\theta}'}_{lwr} & = & \frac{n + 2}{4 ( n + 8 )} \left[ \epsilon   
 + \frac{6 {{\epsilon}'}^2}{n + 8}  \left( \frac{n + 3}{n + 8} + \ln
\frac{3}{2} \right) \right]  \; .
\end{eqnarray}
Clearly the difference between the weak-long-range and short-range regimes 
comes from the difference between $\epsilon$ and ${\epsilon}'$ as given by 
Eq.(15). From the work of \cite{hon90} we know that the LRI fixed point 
becomes instable at $\sigma = {\sigma}_s$. The signal of instability appears 
however at three-loops. Our work shows that already at two-loops a new fixed 
point develops, driving the pure LRI model to the intermediate 
weak-long-range regime. The $\sigma$ dependence of the critical exponent 
${\theta}'$  in this regime is linear and is shown in the curves `d' and `e'
of Figure 3.

We summarise now our results. We studied the short-time critical behaviour
of the Ginzburg-Landau models with  LRI in the  $\epsilon$-expansion  up to
two-loop order.  We observed an initial critical increase for dimensions
smaller than $d_{c}$ and for the interaction range $d/2 <\sigma < d$.
We obtained the universal critical exponents $\theta$ and $\theta^{\prime}$ of
the initial slip as functions of $d$, $n$, and the interaction range
parameter $\sigma$. The limit in which pure LRI is approaching the SRI has 
been also discussed in some detail.

\vskip 1.pc
{\bf Acknowledgements:} The authors are grateful to H. Luo and B. Zheng for 
fruitful discussions, and thank C. Untch for help in using the
computers.

\section*{ Appendix \ The calculation of $Z_{0}$}

\begin{appendix}{}

In order to determine the renormalisation constant $Z_{0}$, we
calculate the two-point function $<s(-q,t)\widetilde
s (q,t^{\prime})>$,
with one leg attached to the initial surface $t'=0$
  \[
  <s(-q,t)\widetilde s (q,0)> =
  \int_{0}^{\infty} dt' <s(-q,t)\widetilde s (q,t^{\prime})>^{(e)}
  \Gamma_{10}^{(i)}(q,t^{\prime}) \; ,
  \]
by using the graphs of Figure 4.

In these diagrams $C_{p}^{(D)}(t,t^{\prime})$  and $G_{p}(t,t^{\prime})$ are
represented by solid lines with and without arrows respectively. The small
circle means that one time argument is set equal to zero. 
The factor  $<s(-q,t)\widetilde s (q,t^{\prime})>^{(e)}$
denotes the contribution to
the two-point function coming only from the equilibrium  part
$C_{p}^{(e)}(t,t^{\prime})$, whereas the residual  factor
$\Gamma_{10}^{(i)}(q,t^{\prime})$ is the sum of the
amplitudes with at least one initial part  $C_{p}^{(i)}(t,t^{\prime})$. 

We write the singular part of $\Gamma_{10}^{(i)}$ at the critical point
$\tau=0$ in the form
\begin{equation}
\Gamma_{10}^{(i)}(q=0,t)=I_{1}+I_{2}+I_{3}+I_{4}+I_{5} \; 
\end{equation}
where $I_j$ with $j=1,2,3,4, 5$ is the contribution of Fig.4($j$). These
contributions are given by
\begin{eqnarray}
I_{1} & = & \delta (t) \; ; \\
I_{2} & = & -\lambda g {n+2 \over 6} \int { d^{d}p \over (2\pi)^{d}}
C_{p}^{(i)}(t,t) \; ; \nonumber \\
I_{3} & = & 2(\lambda g)^{2} \left ( {n+2 \over 6} \right )^{2}
 \int_{0}^{t}dt' \int { d^{d}p \over (2\pi)^{d}}
C_{p}^{(i)}(t,t)
\int { d^{d}p' \over (2\pi)^{d}}
G_{p'}(t,t')C_{p'}^{(D)}(t,t')\; ; \nonumber \\
I_{4} & = & (\lambda g)^{2} {n+2 \over 6}
 \int_{0}^{t}dt' \int { d^{d}p \over (2\pi)^{d}}
{ d^{d}p' \over (2\pi)^{d}}  \nonumber \\
      &   & \mbox{} \times G_{p+p'}(t,t') \left[ 2 \; C_{p}^{(i)}(t,t')
C_{p'}^{(e)}(t,t')+C_{p}^{(i)}(t,t')C_{p'}^{(i)}(t,t') \right] \; ; \\
I_{5} & = & (\lambda g)^{2} \left ( {n+2 \over 6} \right )^{2}
 \int_{0}^{t}dt' \int { d^{d}p \over (2\pi)^{d}}
{ d^{d}p' \over (2\pi)^{d}}
C_{p}^{(i)}(t,t)
C_{p'}^{(i)}(t',t') \; . 
\end{eqnarray}

\vskip 1.pc
\vskip 1.pc

\begin{figure}[h]
\setlength{\unitlength}{0.1in}
\begin{picture}(50,10)
      \put(10,6){\circle{0.5}}
      \put(10,6){\vector(-1,0){5}}
\put(8,3){\makebox(0,0){(1)}}

      \put(26,8){\circle{4}}
      \put(32,6){\line(-1,0){11}}
      \put(30,6){\vector(-1,0){8}}
      \put(31,6){\vector(-1,0){1.5}}
      \put(32,6){\circle{0.5}}
   \put(26,3){\makebox(0,0){(2)}}

      \put(42,8){\circle{4}}
      \put(42,12){\circle{4}}
      \put(40.1,8){\vector(0,-1){0}}
      \put(48,6){\line(-1,0){11}}
      \put(46,6){\vector(-1,0){8}}
      \put(47,6){\vector(-1,0){1.5}}
      \put(48,6){\circle{0.5}}
\put(42,3){\makebox(0,0){(3)}}                  
\end{picture}
\end{figure}

\begin{figure}[h]
\setlength{\unitlength}{0.1in}
\begin{picture}(50,10)
      \put(20,6){\circle{4}}
      \put(26,6){\line(-1,0){11}}
      \put(24,6){\vector(-1,0){8}}
      \put(24.5,6){\vector(-1,0){5}}
      \put(25,6){\vector(-1,0){1.5}}
      \put(26,6){\circle{0.5}}
\put(20,3){\makebox(0,0){(4)}}

      \put(35,8){\circle{4}}
      \put(41,8){\circle{4}}
      \put(41,6){\line(-1,0){11}}
      \put(39.5,6){\vector(-1,0){8}}
      \put(43,6){\vector(-1,0){5}}
      \put(45,6){\vector(-1,0){1.5}}
      \put(45,6){\line(-1,0){11}}
      \put(45,6){\circle{0.5}}
   \put(38,3){\makebox(0,0){(5)}}
   \put(25,0){\makebox(0,0){Fig.4 {\small Diagrams contributing to
   $\Gamma_{10}^{(i)}(q,t)$ up to two loops.}}}

\end{picture}
\end{figure}

\vskip 1.pc 
\vskip 1.pc
By using the formulae
\begin{eqnarray*}
\int_{0}^{\infty} dx  x^{\nu-1} e^{-\mu x} 
& = & \mu^{-\nu} \Gamma(\nu) \; ; \\
\int_{0}^{t} dx x^{\nu - 1}  {( t - x )}^{\mu - 1}
& = & \frac{\Gamma( \nu ) \Gamma ( \mu )}{\Gamma ( \nu + \mu )} \;  
t^{\nu + \mu - 1} \; 
\end{eqnarray*}
valid for ${\bf Re} \nu >0$ and ${\bf Re} \mu >0$, it is not
difficult to obtain
\begin{eqnarray}
I_{2} & = & {n+2 \over 6 \sigma } \lambda g K_{d} \Gamma (1- \epsilon /
\sigma) (2\lambda t)^{-1+\epsilon/ \sigma} \; ; \\
I_{3}& = & -\left ( {n+2 \over 3} \right )^{2}\lambda (g K_{d})^{2}
{\Gamma^{2} (1-\epsilon / \sigma) \over 2\sigma \epsilon} \; \nonumber \\
      &   & \mbox{} \times \left[ {{\Gamma}^{2} (1 + \epsilon / \sigma) \over
\Gamma (1 + 2 \epsilon / \sigma) }-{1 \over 2} \right]
(2\lambda t)^{-1+2\epsilon/ \sigma} \; ;  \\
I_{5}& = & \left ( {n+2 \over 6} \right )^{2}
\lambda (g K_{d})^{2}
{{\Gamma}^{2} (1 - \epsilon / \sigma) \over 2 \sigma \epsilon } 
(2\lambda t)^{-1+2\epsilon/ \sigma}\; .
\end{eqnarray}
By integrating in (21) over $t'$ and over the length of $p'$, one finds
\[
I_{4}=
{n+2 \over 6\sigma}\lambda (g K_{d})^{2}
\Gamma (1 - 2 \epsilon / \sigma)
\left[ -I_{4}^{(1)}+I_{4}^{(2)}+O(\epsilon) \right] 
(2\lambda t)^{-1+2\epsilon/ \sigma} \; 
\]
where 
\begin{eqnarray*}
I_{4}^{(1)} & \equiv & K_{d}^{-1}
\int{d^{d} x \over (2 \pi)^{d}} {1\over x^{\sigma} ({\bf e+x})^{\sigma} }
\; ; \\
I_{4}^{(2)} & \equiv & K_{d}^{-1}\int
{d^{d} x \over (2 \pi)^{d}} {1\over x^{\sigma} ({\bf e+x})^{\sigma
}( 1+x^{\sigma})} \; ,
\end{eqnarray*}
with $\bf{e}$ the unit vector along the $d$-axis.
The first integral is easily done with the result
\begin{eqnarray*}
I_{4}^{(1)}
& = & {K_{d}^{-1} \over 2^{d} \pi^{d/2}} 
{{\Gamma}^2 \left( \displaystyle{\frac{d - \sigma}{2}} \right) 
\Gamma \left( \displaystyle{\frac{2 \sigma - d}{2}} \right) 
\over {\Gamma}^2 \left( \displaystyle{\frac{\sigma}{2}} \right) 
\Gamma (d - \sigma )} \; \\
& = & {1 \over \epsilon} + \frac{D_{\sigma}}{2}  + O(\epsilon) \; .
\end{eqnarray*}
In order to carry out $I_{4}^{(2)}$,
one can use the following expansion of $1/({\bf e} + {\bf x})^{\sigma}$:
\[
(x^{2} + 2 {\bf x}\cdot{\bf e} + 1)^{-\sigma/2}
=\left [ \max (x,1) \right  ]^{- \sigma}
\sum\limits_{n=0}^{\infty}\left [ \min (x,1/x) \right  ]^{n}
(-1)^{n}c_{n}^{\sigma/2}({\bf \hat{x}}\cdot {\bf e})\;
\]
where  ${\bf \hat{x}}$ stands for the unit vector of ${\bf x}$ and
$c_{n}^{\sigma/2}({\bf \hat{x}}\cdot {\bf e})$ are Gegenbauer polynomials.
This leads to
\[
\int d\hat{x} ({\bf x+e})^{-\sigma}
= \left \{
\begin{array}{cc}
1 \quad \quad  x\le 1    \\
x^{-\sigma} \quad  \quad  x \ge 1
\end{array}
\right.
\]
which can be then used to find the leading contribution to $I_4^{(2)}$
\[
I_{4}^{(2)}
={2 \over \sigma}\ln 2 +O({\epsilon}).
\]
Finally, we get
\begin{equation}
I_{4}
={n+2 \over 6\sigma}\lambda (g K_{d})^{2}
\Gamma \left( 1 - {2\epsilon \over \sigma} \right) 
\left[ - \frac{1}{\epsilon} + {2 \over \sigma} \ln 2
- {D_{\sigma} \over 2} + O(\epsilon) \right]
(2\lambda t)^{2 \epsilon / \sigma - 1} \; .
\end{equation}
The substitution of Eqs.(20),(23-26) in (17) leads to 
an explicit expression for $\Gamma_{10}^{(i)}(q=0,t)$ up to terms of order 
$\epsilon$. 

We renormalise now according to (4) the (bare) quantities entering this
expression. For the fields $s$, $\widetilde s$ and 
the coupling constant $g$ a one-loop renormalisation will be sufficient. 
All the necessary information is available in Eqs.(5-7). The
residual singularity is then removed by requiring \cite{s1}
  \[
  Z_{0}^{-1/2}
  \int_{0}^{\infty}dt e^{-i\omega t}
  \Gamma_{10}^{(i)}(q=0,t)_{b} = \mbox{finite for $\epsilon \rightarrow
0$} \; .  \]
Here the subscript $b$ denotes the expression of $\Gamma_{10}^{(i)}$ obtained 
above in which only bare quantities appear.
From this condition we compute $Z_{0}$ as given by Eq.(9).

\end{appendix}


\begin{figure}[p]\centering
\epsfysize=12cm
\epsfclipoff
\fboxsep=0pt
\setlength{\unitlength}{1cm}
\begin{picture}(13.6,12)(0,0)
\put(0,0){{\epsffile{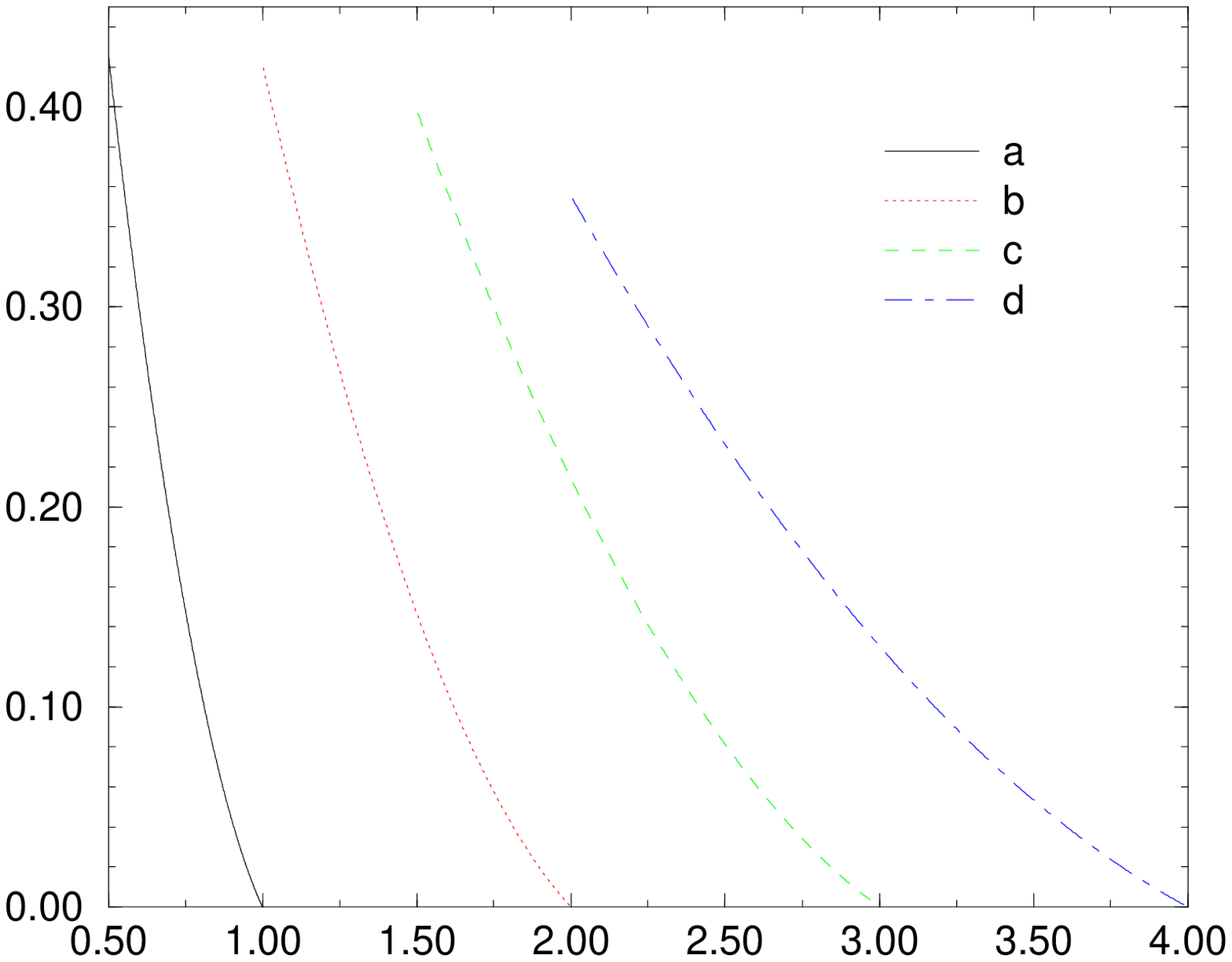}}}
\put(1.5,9){\makebox(0,0){$\theta^{\prime}$}}
\put(11,1){\makebox(0,0){$d$}}
\end{picture}
\caption{The exponent $\theta^{\prime}$ for $n=1$ is plotted versus $d$.
Curves `a', `b', `c', and `d' correspond to 
$\sigma=1/2, \ 1, \ 3/2$ and $2$ 
respectively. }
\label{f1}
\end{figure}

\begin{figure}[p]\centering
\epsfysize=12cm
\epsfclipoff
\fboxsep=0pt
\setlength{\unitlength}{1cm}
\begin{picture}(13.6,12.)(0,0)
\put(0,0){{\epsffile{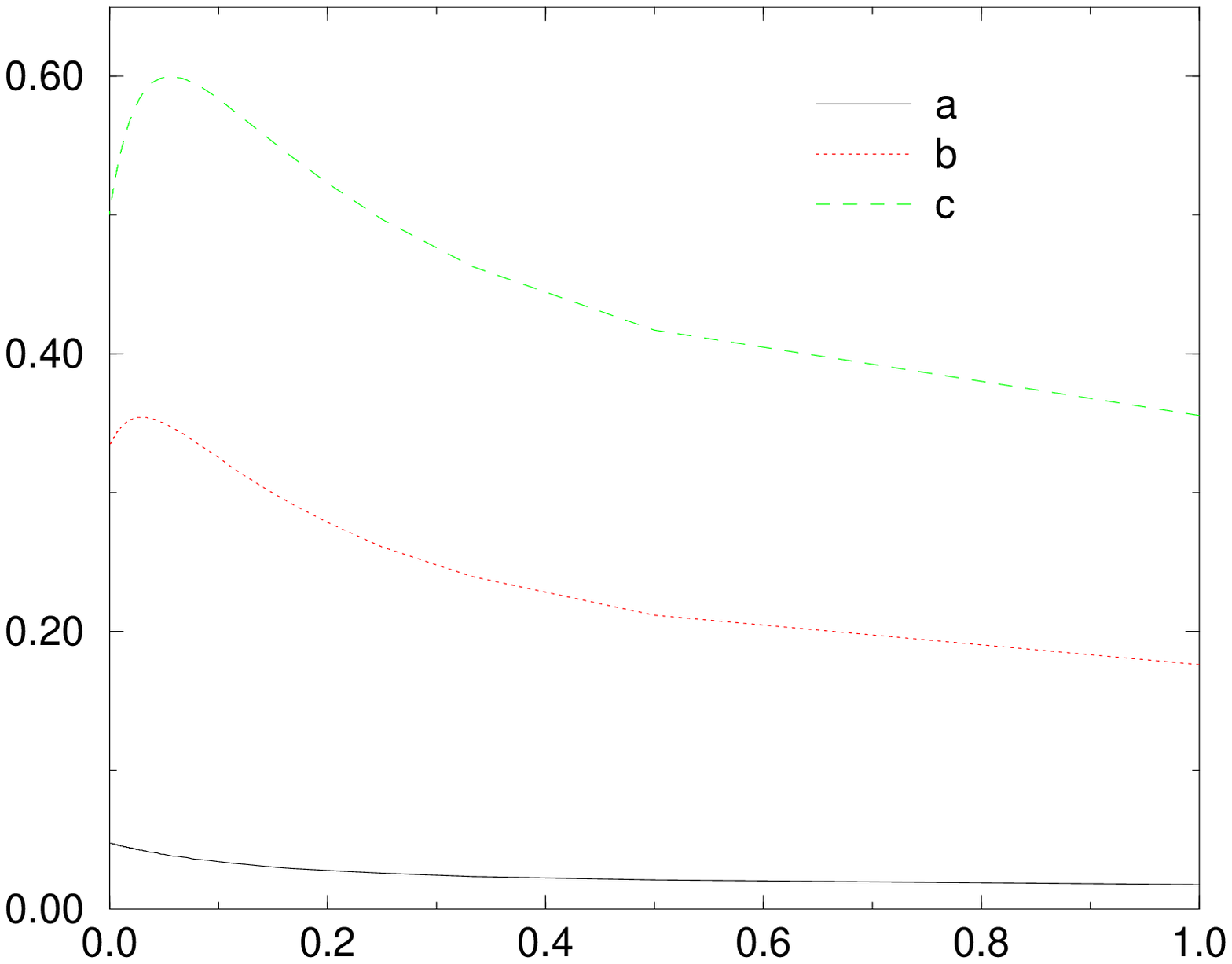}}}
\put(1.5,9.5){\makebox(0,0){$\theta^{\prime}$}}
\put(11,1){\makebox(0,0){$1/n$}}
\end{picture}
\caption{
The exponent $\theta^{\prime}$ for $d=2$ is plotted versus $1/n$.
The curves `a', `b', and `c' correspond  to $\sigma=1.05$, $3/2$, and
$2$ respectively.
}
\label{f2}
\end{figure}

\begin{figure}[p]\centering
\epsfysize=12cm
\epsfclipoff
\fboxsep=0pt
\setlength{\unitlength}{1cm}
\begin{picture}(13.6,12.)(0,0)
\put(0,0){{\epsffile{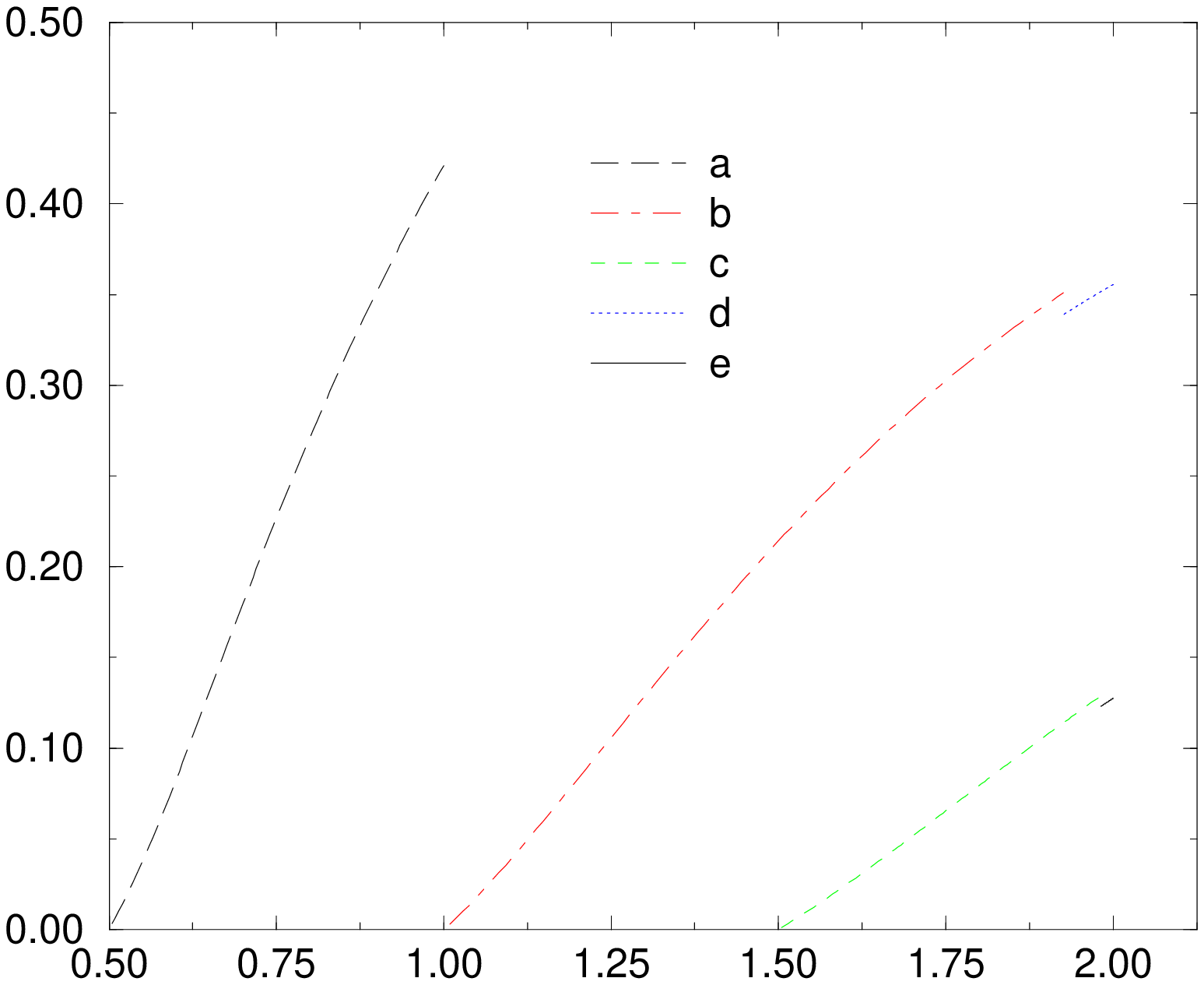}}}
\put(1.5,10){\makebox(0,0){$\theta^{\prime}$}}
\put(11,1){\makebox(0,0){$\sigma$}}
\end{picture}
\caption{The exponent $\theta^{\prime}$ for $n=1$ is plotted versus $\sigma$.
The curve `a'  corresponds to $d=1$.
The curves `b' and `c' are drawn for $\sigma \leq \sigma_{s}$
at $d=2$ and $d=3$, respectively; the curves `d' and `e'
are based on Eq.(18) and describe the behaviour in the region
$\sigma \geq {\sigma}_{s}$.
}
\label{f3}
\end{figure}

\end{document}